%% file: preprint/main_preprint.tex
\documentclass[10pt, a4paper, twocolumn]{article}
\usepackage{hyperref}
\usepackage{url}
\usepackage{booktabs}
\usepackage[normalem]{ulem}

\input{math_commands.tex}

\newcommand{\predError}{\epsilon}
\newcommand{\resOutput}{\mathbf{o}}
\newcommand{\delay}{\tau_{\text{d}}}
\newcommand{\vel}{\mathbf{v}}
\newcommand{\scalarvel}{v}
\newcommand{\partpos}{\mathbf{r}}
\newcommand{\rampLength}{\Gamma}
\newcommand{\laserInt}{I}
\newcommand{\Izero}{I_{\textnormal{0}}}
\newcommand{\tarDist}{\mathbf{d}_{\textnormal{t}}}
\newcommand{\tarDisti}{\mathbf{d}_{\textnormal{t,i}}}
\newcommand{\abstarDist}{D_{\textnormal{target}}}
\newcommand{\angMom}{h}
\newcommand{\tarpos}{\mathbf{q}}
\newcommand{\tarMotAng}{\beta}
\newcommand{\gridDist}{L}
\newcommand{\tRelax}{\tau_{r}}
\newcommand{\meanAbsVel}{\langle \scalarvel \rangle}
\newcommand{\moveVar}[1]{\sigma^2_{10}[#1]}
\newcommand{\signal}{\mathbf{u}}
\newcommand{\resState}{\mathbf{s}}
\newcommand{\readout}{\mathbf{x}}
\newcommand{\predLen}{\Delta t}
\newcommand{\coeffDet}{\mathrm{R}2}
\newcommand{\driverFract}{f_{\mathrm{driver}}}
\newcommand{\Nhist}{N}
\newcommand{\readoutLayer}{\mathbb{W}_\mathrm{out}}
\newcommand{\timeComp}{f_c}
\newcommand{\nrmse}{N\!R\!M\!S\!E}
\newcommand{\fscore}{F_{1}}

\usepackage[margin=0.75in]{geometry}
\usepackage{titlesec}
\titleformat{\section}{\large\bfseries}{\thesection}{1em}{}
\titlespacing*{\section}{0pt}{2ex}{0.5ex}
\titleformat{\subsection}{\normalsize\bfseries}{\thesubsection}{1em}{}
\titlespacing*{\subsection}{0pt}{1ex}{0ex}
\usepackage{abstract}

\usepackage{graphicx} 
\usepackage[font={footnotesize}, labelfont={bf}]{caption}
\usepackage{float}
\newcounter{sifig}

\newfloat{suppfig}{htbp}{losf}
\makeatletter
\renewcommand{\fps@suppfig}{htbp}

\makeatother

\usepackage[
    style=nature,
    sorting=none,
    url=false,
    isbn=false]{biblatex}
\AtEveryBibitem{
  \clearlist{language}
  \clearfield{language}
}
\title{Reservoir computing from collective dynamics of active colloidal oscillators}
\date{January 2026}
\usepackage{authblk}
\author[1, 2]{Veit-Lorenz Heuthe}
\author[1]{Lukas Seemann}
\author[3]{Samuel Tovey}
\author[1, 2]{Clemens Bechinger\thanks{Corresponding author, email: clemens.bechinger@uni.kn}}
\affil[1]{\small{Fachbereich Physik, Universität Konstanz, Universitätsstr. 10, 78464 Konstanz, Germany}}
\affil[2]{\small{Centre for the Advanced Study of Collective Behaviour, Universität Konstanz, Universitätsstr. 10, 78464 Konstanz, Germany}}
\affil[3]{\small{Institute for Computational Physics, University of Stuttgart, Allmandring 3, 70569 Stuttgart, Germany}}

\begin{document}
\twocolumn[\begin{@twocolumnfalse}
\maketitle

\begin{abstract}
\bf{Physical reservoir computing is a computational framework that offers an energy- and computation-efficient alternative to conventional training of neural networks. In reservoir computing, input signals are mapped into the high-dimensional dynamics of a nonlinear system, and only a simple readout layer is trained. In most physical implementations, the interactions that give rise to the dynamics cannot be tuned directly and high dimensionality is typically achieved through time-multiplexing, which can limit flexibility and efficiency. Here we introduce a reservoir composed of hundreds of hydrodynamically coupled active colloidal oscillators forming a fully parallel physical reservoir and whose coupling strength and fading-memory time can be tuned in situ. The collective dynamics of the active oscillators allow accurate predictions of chaotic time series from single reservoir readouts without time-multiplexing. We further demonstrate real-time detection of subtle hidden anomalies that preserve all instantaneous statistical properties of the signal yet disrupt its underlying temporal correlations. These results establish interacting active colloids as a reconfigurable platform for physical computation and edge-integrated intelligent sensing for model-free detection of irregularities in complex time signals.}
\end{abstract}
\vspace{2em}
\end{@twocolumnfalse}]
\saythanks

\begin{figure*}[t]
    \centering
    \includegraphics[width=1\linewidth]{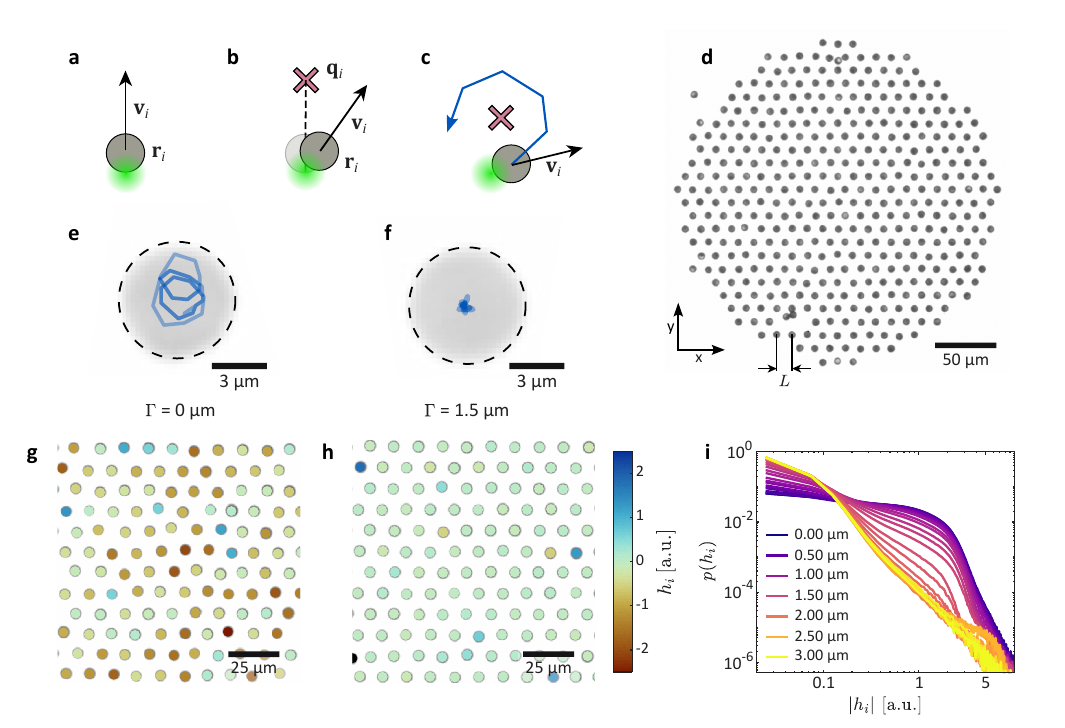}
    \caption{
        \textbf{Active colloidal oscillators.}
        \textbf{a} A colloidal particle (grey) with position $\partpos_i$ is activated by a focused laser beam (green), inducing self-propulsion with velocity $\vel_i$.
        \textbf{b} Active colloid steered toward a target position $\tarpos_i$ (red). The propulsion direction $\vel$ (solid arrow) deviates from the target direction (dashed line) due to a time delay between particle detection and laser positioning.
        \textbf{c} This misalignment induces an orbiting motion of the active colloid around the target position (blue arrow).
        \textbf{d} Experimental snapshot of 400 ACOs with target positions arranged on a hexagonal lattice of spacing $\gridDist$.
        \textbf{e,f} Trajectories (blue) of a single particle for damping parameters $\rampLength$~=~0 and $1.5\mathrm{\mu m}$. The gray area corresponds to the optical image of the particle (dashed circle).
        \textbf{g,h} Section of the oscillator array shown in \textbf{D}, with ACOs colored by their angular momentum $\angMom_i$ for $\rampLength = 0$ and $1.5\mathrm{\mu m}$.
        \textbf{i} Experimentally measured probability distributions of $\left| \angMom_i \right| = \left| \vel_i \times (\partpos_i - \tarpos_i) \right|$ for different values of $\rampLength$.
            }
    \label{fig:1}
\end{figure*}

\section*{Introduction}
The rapid expansion of artificial intelligence has intensified the demand for energy-efficient computing architectures. Contemporary machine-learning approaches, particularly deep neural networks, rely on optimizing millions of parameters and therefore require substantial computational and energy costs. As data volumes and task complexity continue to grow, alternative paradigms that perform computation directly within physical substrates (in materia) are attracting increasing interest. Reservoir computing (RC) offers such a route: instead of training all internal connections of a neural network, RC exploits the intrinsic dynamics of a nonlinear system—--the reservoir—--to transform input signals into a high-dimensional state space, while learning is confined to a simple linear readout. Physical reservoirs naturally provide nonlinearity and intrinsic memory, enabling compute-in-memory architectures that bypass the von Neumann bottleneck~\cite{kumar_dynamical_2022, duan_memristorbased_2024}.

RC has been explored on a wide range of physical platforms, including optical feedback networks~\cite{brunner_parallel_2013, rafayelyan_large-scale_2020}, memristor arrays~\cite{du_reservoir_2017}, magnetic textures~\cite{beneke_gesture_2024}, quantum devices~\cite{martinez-pena_input-dependence_2025, mujal_opportunities_2021, tovey_generating_2025}, and chemical or electrochemical systems~\cite{cucchi_reservoir_2021, baltussen_chemical_2024}. These systems are particularly well suited for processing temporal signals with strong dependence on their history, enabling applications such as forecasting chaotic dynamics, interpolating missing data, and detecting anomalies in fields ranging from physiology and seismology to climate science and finance~\cite{lee_physical_2025, noor_machine_2025, cardoso_echo_2024, walleshauser_predicting_2022, santos_reservoir_2025, yao_training_2015, nadiga_reservoir_2021}. However, in most existing physical reservoirs, internal interactions are fixed or only weakly tunable, and high effective dimensionality is often achieved through time-multiplexing rather than by exploiting intrinsically rich reservoir dynamics~\cite{cucchi_hands-reservoir_2022}. This limits true parallel processing, restricts access to optimal regimes that balance nonlinearity and fading memory, and hampers systematic optimization, since modifying interaction strength or network topology typically requires the fabrication of new devices.

These limitations motivate the search for physical reservoirs whose computational richness arises intrinsically from genuine many-body dynamics and whose interactions can be reconfigured in situ. Active matter, i.e., ensembles of self-driven colloidal particles with highly nonlinear and long-range interactions—provides a natural realization of such systems~\cite{gaimann_robustly_2025, lymburn_reservoir_2021, gaimann_optimal_2025}. Although recent work has shown that single synthetic active particles with delayed feedback can serve as nonlinear physical reservoir nodes~\cite{wang_harnessing_2024}, it remains an open challenge to realize reconfigurable reservoirs in which computational function emerges from collective, spatially extended dynamics of interacting units. Harnessing such many-body dynamics promises intrinsically high dimensionality, true parallelism, and computational capabilities that are inaccessible in single-node or time-multiplexed architectures.

Here we demonstrate a physical reservoir realized as a hexagonal lattice of hundreds of hydrodynamically coupled active colloidal oscillators (ACOs). The platform offers full experimental control over coupling strength, lattice geometry, and signal injection, enabling systematic exploration and optimization of reservoir performance. Computation is encoded in the collective spatiotemporal dynamics of the array and proceeds in parallel across hundreds of interacting degrees of freedom, without reliance on time-multiplexing or virtual node construction. These collective dynamics enable accurate forecasting of chaotic time series and, crucially, real-time detection of hidden anomalies within chaotic signals. Together, our results establish colloidal active-matter reservoirs as a versatile experimental platform for edge-integrated intelligent sensing and processing complex temporal data.

\section*{Active colloidal oscillator arrays}
\begin{figure*}[t]
    \centering
    \includegraphics[width=1\linewidth]{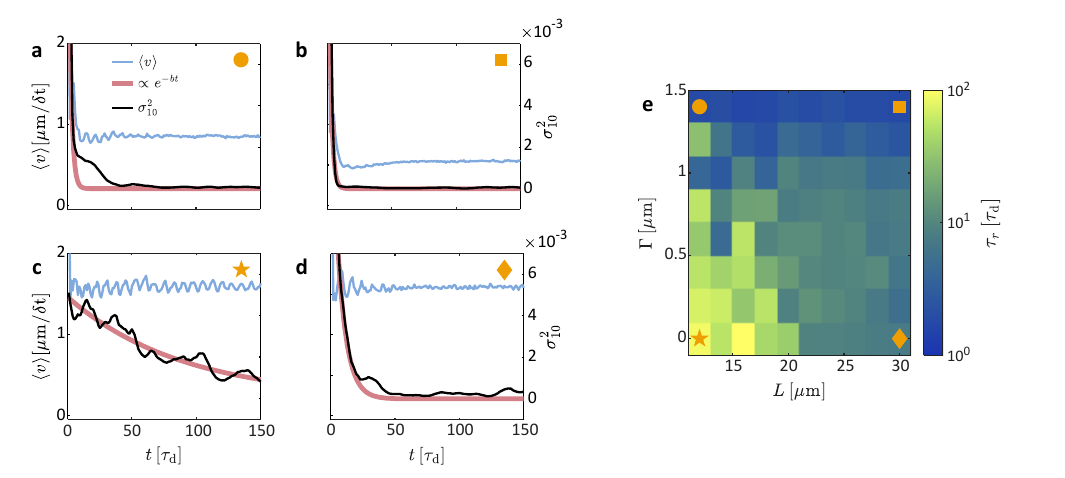}
    \caption{
        \textbf{Response of reservoir of 400 ACOs to an input signal.}
        \textbf{a-d} Relaxation of the mean absolute velocity $\meanAbsVel$ (blue) after excitation with a $4\,\mathrm{\delay}$ square pulse for different parameters 
        (A: $\rampLength = 1.4\,\mathrm{\mu m}, \gridDist = 12\,\mathrm{\mu m}$, 
        B: $\rampLength = 1.4\,\mathrm{\mu m}, \gridDist = 30\,\mathrm{\mu m}$, 
        C: $\rampLength = 0\,\mathrm{\mu m}, \gridDist = 12\,\mathrm{\mu m}$, 
        D: $\rampLength = 0\,\mathrm{\mu m}, \gridDist = 30\,\mathrm{\mu m}$).
        The black and red curves are the running variance $\moveVar{\meanAbsVel}$ with a window length of 10 $\delay$ and an exponential fit to $\moveVar{\meanAbsVel}$, respectively.
        \textbf{e} Color map of the measured relaxation times $\tRelax$ for different combinations of $\rampLength$ and $\gridDist$. Orange symbols indicate the parameter combinations illustrated in A-D. See Fig.~S1 for corresponding simulation results.
            }
    \label{fig:2}
\end{figure*}

As an experimental realization of a reservoir computing substrate, we employ periodic arrays of hydrodynamically coupled active colloidal oscillators (ACOs)~\cite{heuthe_tunable_2025}. Each ACO is formed by steering a self-propelled colloidal particle toward a fixed target position $\tarpos_i$ using a feedback-controlled scanning laser. The propulsion velocity $\vel$, and thus the oscillator dynamics, can be continuously tuned by adjusting the laser intensity and steering. A finite feedback delay $\delay$ between particle detection and laser repositioning introduces a controlled misalignment in the propulsion direction, generating stable orbital motion around the target (Figs.~\ref{fig:1}a-c and Supplementary Movie~1). This setup allows simultaneous control of several hundred oscillators (see Methods).
Arranging the oscillators on a hexagonal lattice couples their motion via the flow fields within the liquid and generates complex collective dynamics~\cite{heuthe_tunable_2025, hu_hovering_2024, kotar_hydrodynamic_2010} (Fig.~\ref{fig:1}d). In this configuration, the ACOs serve as the nodes of a physical reservoir, where hydrodynamic interactions mediate information processing and shape the network’s nonlinear dynamics~\cite{csaba_coupled_2020}. The coupling strength, and thus the reservoir’s functionality, can be continuously tuned by adjusting the lattice spacing $\gridDist$, since hydrodynamic interactions decay with distance~\cite{heuthe_tunable_2025}.
To control the sensitivity of ACOs to external inputs, we tune their effective damping: when the displacement of the colloid's position $\partpos_i$ from the target position $|\partpos_i - \tarpos_i|$ falls below a threshold $\rampLength$, the propulsion speed is linearly reduced. This damping regulates the oscillation amplitude and is quantified via each oscillator’s specific angular momentum $\angMom_i = \vel_i \times (\partpos_i - \tarpos_i)$ (Figs.~\ref{fig:1}e-f). As shown in Figs.~\ref{fig:1}g-h, increasing $\rampLength$ suppresses noise-dominated fluctuations and produces more regular dynamics, continuously narrowing the distribution of $\angMom$ (Fig.~\ref{fig:1}i) (Methods).
To quantify how long signals persist within the reservoir, we measured the response of the ACO array after excitation with a single rectangular pulse, during which we displace all targets along the $x$-direction. Strong ACO-coupling at small $\gridDist$ leads to a collective motion displayed by pronounced oscillations of the mean particle velocity $\langle \scalarvel \rangle$, which weaken with increasing spacing and with added damping (Figs.~\ref{fig:2}a-d). By tuning $\rampLength$, we can therefore adjust both the transient response and the long-term relaxation. Fitting the decay yields relaxation times $\tRelax$  that vary by more than an order of magnitude, demonstrating how damping and spacing jointly govern the reservoir’s memory (Fig.~\ref{fig:2}e).

\section*{ACO-based reservoir computing}
\begin{figure*}[t]
    \centering
    \includegraphics[width=1\linewidth]{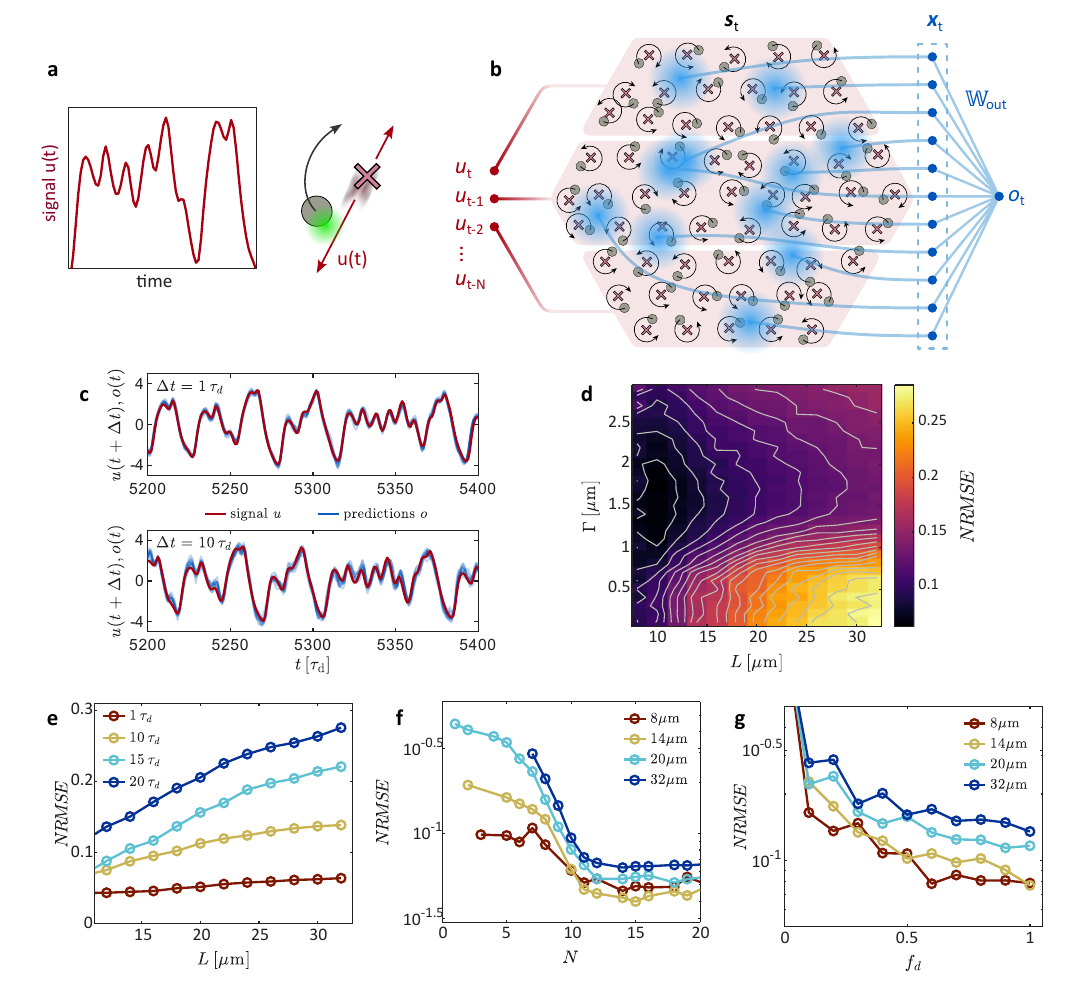}
    \caption{
        \textbf{Reservoir computing with arrays of coupled ACOs.}
        \textbf{a} Coupling of the input signal (red) to the active colloidal oscillators (ACOs). The signal modulates each oscillator’s target position $\tarpos$, which is translated along a line with a fixed but randomly chosen orientation.
        \textbf{b} Working principle of the reservoir. The signal $\signal(t)$ drives all ACOs within the reservoir which is divided into $\Nhist$ rows with a one-step delay between rows, which propagates the signal through the ACO array. 
        From the full reservoir state $\resState(t) = \left\{ \partpos_1(t), \vel_1(t), ... \, \partpos_i(t), \vel_i(t) \right\}$, we construct readout vectors $\readout(t)$ using Gaussian kernels (blue; see Methods). One linear readout layer $\readoutLayer$ produces the output $\resOutput(t)$ (e.g. a prediction of the signal $\signal(t + \predLen)$).
        \textbf{c} Representative segment of a chaotic Mackey–Glass (MG) signal $\signal(t + \predLen)$ (red) together with output $\resOutput(t)$ (blue) of reservoirs trained to predict the signal with forecasting horizons $\predLen = 1$ (top) and $\predLen = 10$ (bottom) from the validation split (i.e. not part of the training data) from experiments.
        \textbf{d} Heat map of the average prediction performance $\nrmse$ for $\predLen = 10$ MG forecasting ($\Nhist = 10$) in simulations as a function of ACO spacing $\gridDist$ and damping parameter $\rampLength$.
        \textbf{e} Average prediction performance $\nrmse$ for ten step MG forecasting in simulations as a function of ACO separation $\gridDist$ in simulations ($\Nhist = 10$) for different prediction horizons $\predLen$.
        \textbf{f} Change of average error $\nrmse$ for ten step MG forecasting with increasing number of injected signal states $\Nhist$ for different ACO separations $\gridDist$ in simulation. See Fig.~S3 for a larger range of $\Nhist$.
        \textbf{g} Average error for ten step MG forecasting as a function of the fraction of driven ACOs $\driverFract$ and for different $\gridDist$ in simulations ($\Nhist = 10$).
        }
    \label{fig:3}
\end{figure*}

The single-pulse experiments demonstrate that brief shifts of the target positions induce complex and highly non-linear relaxation dynamics that are captured by the particles’ velocity response. This establishes not only controlled target displacements as a suitable and effective way to inject information into the reservoir, but also particle velocities as a meaningful readout of its internal state. Building on this, we now drive the reservoir by varying the target positions $\tarpos_i(t)$ of all oscillators according to an external input signal (see Fig.~\ref{fig:3}a and Supplementary Movie~1). Each target moves along a fixed, randomly chosen direction to ensure strong hydrodynamic coupling across the array. By dividing the array into $\Nhist$ rows and applying the input with controlled delays between rows, the excitation propagates through the lattice (Fig.~\ref{fig:3}b), allowing signal values from different times to interact and thereby enhancing both memory and computational richness.

The reservoir’s response to the input yields a time series of reservoir states $\resState(t)$ comprised of the positions $\partpos$ and velocities $\vel$ of all oscillators  $\resState(t) = \left\{ \partpos_1(t), \vel_1(t), ... \, \partpos_i(t), \vel_i(t) \right\}$. To mitigate Brownian noise and temporary defect ACOs, we compute the output $\resOutput$ from a readout $\readout$, which is constructed from 1000 Gaussian kernels following Ref.~\cite{lymburn_reservoir_2021} (blue in Fig.~\ref{fig:3}b, see Methods and Supplementary Movie~2). Each kernel computes a locally averaged particle density and velocity using a weighting function $\exp\left[-(\partpos-\partpos_i)^2/2\sigma^2\right]$ centered at a randomly chosen position. The kernel width and number were optimized for performance (Fig.~S2). Finally, the reservoir output is obtained via
\begin{equation}
    \resOutput(t) = \readoutLayer \readout(t)^\top.
\end{equation} 
with the weights of $\readoutLayer$ trained separately for each specific task.
\clearpage

\section*{Chaotic time series forecasting}
\begin{figure*}[t]
    \centering
    \includegraphics[width=1\linewidth]{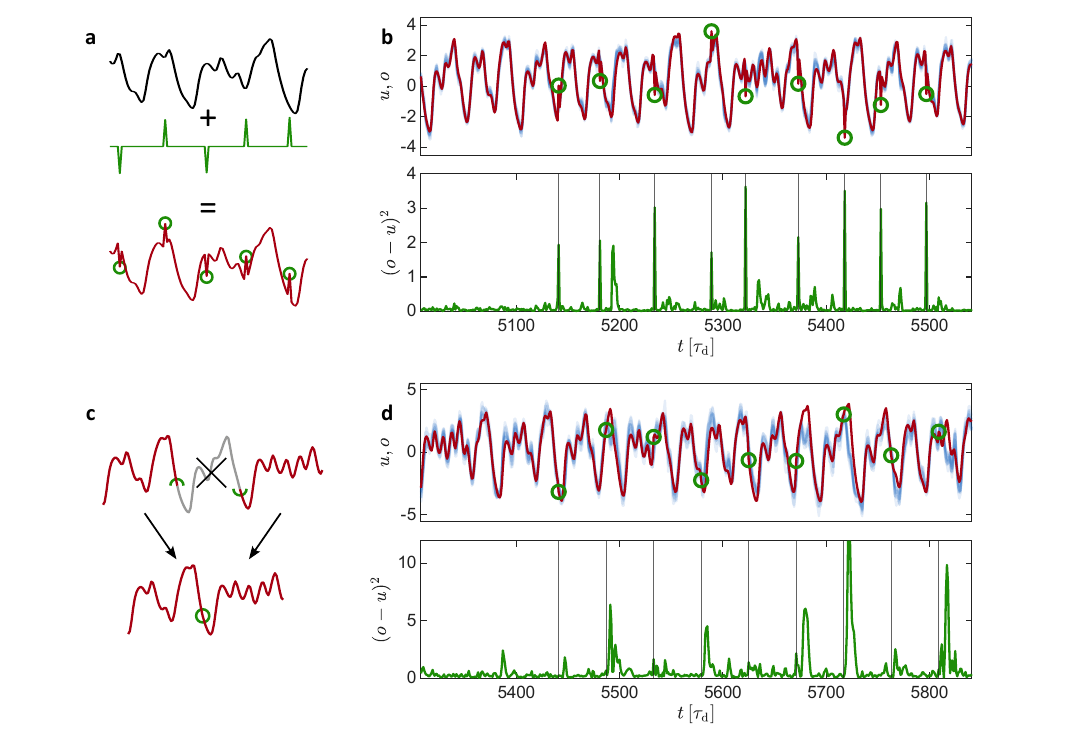}
    \caption{
        \textbf{Detecting hidden anomalies via prediction errors.}
        \textbf{a} Introduction of spiking anomalies to the Mackey–Glass (MG) series (black) by adding a sparse spiking signal (green), resulting in local outliers (red; spikes highlighted by green circles).
        \textbf{b} Top: MG signal $\signal$ containing spiking anomalies (red) together with reservoir predictions $\resOutput$ (light blue). Bottom: Squared prediction error $\left(\resOutput - \signal\right)^2$ over time, showing pronounced peaks at the anomaly locations (vertical lines).
        \textbf{c} Introduction of hidden anomalies by replacing short signal segments with intervals matching both instantaneous value and slope, preserving the visible continuity of the time series while altering its hidden memory structure.
        \textbf{d} Top: MG signal with hidden anomalies (red; highlighted by green circles) and corresponding predictions $\resOutput$ (light blue). Bottom: Squared prediction error $\left(\resOutput - \signal\right)^2$, revealing clear peaks at the hidden anomaly times (vertical lines).
        }
    \label{fig:4}
\end{figure*}
A particular strength of reservoir computing is its capacity to forecast chaotic signals through its fading memory. We demonstrate this by predicting a chaotic Mackey–Glass (MG) time series, a standard benchmark for nonlinear prediction~\cite{wringe_reservoir_2025} (for details see Methods). In such forecasting tasks, the reservoir output $\resOutput(t)$ is trained to predict the signal value $\signal(t+\predLen)$ at a future time $t+\predLen$ by optimizing the weights of the readout layer $\readoutLayer$ to minimize the prediction error $\predError = \left\langle \left( \resOutput - \signal \right) \right\rangle_t$.

In our experiments, we applied an MG input signal $\signal(t)$ for 6700 time steps to an array of 400 active colloidal oscillators and recorded the corresponding series of readouts $\readout$ (see Supplementary Movie~2). The first 80\% of these data (training split) were used to train the linear readout layer $\readoutLayer$ and the forecasting performance was then evaluated for the remaining 20\% of the time steps (validation split, see Methods). Figure~\ref{fig:3}c shows a time segment of the output $\resOutput(t)$ compared to the input signal $\signal(t+\predLen)$ for a reservoir with $L = 14\,\mathrm{\mu m}$ and $\Gamma = 1.5\,\mathrm{\mu m}$ and for forecasting intervals $\predLen = 1$ and $\predLen = 10$, respectively. Despite minor variations between repeated experimental runs, the output values closely follow the original MG signal demonstrating the capability of our system as a physical reservoir.

The predictive performance is quantified by the normalized root-mean-squared error $\nrmse = \sqrt{\predError / \sigma^2}$ between the forecast and true signal across the entire validation split, with $\sigma^2$ denoting the signal variance. This metric compares the prediction error to the intrinsic fluctuations of the input signal and yields $\nrmse = 0$ for perfect forecasts. Without any predictive power, a linear readout optimized with ridge-regression would result in $\nrmse \approx 1$, corresponding to simply returning the mean value of the signal in every time step. For the example predictions shown in Fig.~\ref{fig:3}c ($L = 14~\mu\mathrm{m}$, $\Gamma = 1.5~\mu\mathrm{m}$, $\Nhist = 10$, $\predLen = 1$), we obtain $\nrmse \approx 0.1$. Notably, this forecasting accuracy is achieved without any time-multiplexing---a feature that distinguishes our platform from nearly all existing physical reservoirs, including photonic, memristive, and spintronic systems~\cite{wang_harnessing_2024, lee_task-adaptive_2024, lu_memristor-based_2024, canaday_rapid_2018}. The dimensionality here arises solely from the simultaneous nonlinear dynamics of hundreds of interacting oscillators. Notably, robust forecasting is maintained even though a few ACOs occasionally deviate from ideal behavior, including brief periods of incomplete response to the laser-steering signal or the formation of short-lived phoretic clusters (Fig.~\ref{fig:1}d and Supplementary Movie~1). The emergence of reliable prediction under these minimally controlled conditions reveals the inherent computational capacity of interacting active-matter systems.

As discussed above, both the damping parameter $\rampLength$ and the oscillator spacing $\gridDist$ strongly shape the array’s relaxation dynamics and thus its forecasting performance. To explore this parameter space without prohibitive measurement times, we performed corresponding numerical simulations. The dynamics of ACOs are well captured by Langevin simulations of active Brownian particles whose hydrodynamic interactions are modeled as a superposition of lubrication forces and squirmer interactions~\cite{heuthe_data_2025} (Methods). Figure~\ref{fig:3}d shows $\nrmse$ for one-step MG-series prediction in simulated reservoirs as a function of $\rampLength$ and $\gridDist$. The prediction accuracy decreases monotonically (i.e., $\nrmse$ increases) with increasing $\gridDist$, confirming that distance-dependent hydrodynamic coupling is essential for reservoir performance. Depending on $\rampLength$ and $\gridDist$, the forecasting error $\nrmse$ varies by more than a factor of three, underscoring the need to finely tune the dynamics of the individual ACOs that form the reservoir nodes. This sensitivity becomes even more pronounced for longer forecasting times $\predLen$ (Fig.~\ref{fig:3}e): as $\predLen$ increases, the slope of $\nrmse(\gridDist)$ steepens and the saturation point shifts to smaller separations, highlighting the growing importance of strong hydrodynamic coupling for more demanding prediction tasks. The high sensitivity of the prediction performance on the experimental parameters illustrates the value of tunability in physical reservoirs and highlights a key advantage of active-matter reservoirs over platforms with fixed interaction networks.

A key advantage of our locally addressable and spatially extended reservoir is that different regions can be driven independently. Rather than relying on serial time-multiplexing within a single node, we exploit spatial selectivity by injecting time-shifted copies of the input signal into distinct subsets of ACOs. This produces controlled spatiotemporal excitation patterns that enhance information propagation across the array and enable interactions between signal components introduced at different times. As shown in Fig.~\ref{fig:3}f, increasing the number of simultaneous input channels $\Nhist$ from 1 to 10 substantially improves the accuracy of the forecast, while higher values degrade performance. This indicates that excessive input complexity can overwhelm the reservoir dynamics.
In experiments, we find a similar dependence of $\nrmse$ on $\Nhist$ (see Fig.~S3c).
The results reported here assume coherent driving of all ACOs; however, in practical implementations uniform input coupling may not always be achievable. To assess the robustness of the reservoir under such conditions, we therefore applied the input signal only to a randomly selected fraction $\driverFract$ of ACOs. As shown in Fig.~\ref{fig:3}g, the reservoir maintains high predictive accuracy even when only 20 \% of ACOs are driven; however, at larger separations $\gridDist$ performance decreases and higher driving fractions become necessary. These findings demonstrate that hydrodynamic interactions efficiently distribute information throughout the array, enabling rich collective dynamics even under partial driving.
Figure~S3a shows the dependence of $\nrmse$ on $\driverFract$ in experiments, which agrees well with the simulation results.
In addition, we validated the reservoir’s forecast performance on another established benchmark, the Lorenz attractor, which arises from a set of nonlinear differential equations originally introduced to describe fluid convection~\cite{wringe_reservoir_2025, lorenz_deterministic_1963} (for details, see Methods and Fig.~S4). We find that for the Lorenz attractor forecasting task, the qualitative dependence of the $\nrmse$ on $\gridDist$ and $\rampLength$ agrees with Fig.~\ref{fig:3}d, showing that these results are independent of the specific time series.

\section*{Hidden anomaly detection}
The forecasting capabilities demonstrated above are particularly valuable for detecting anomalies in real-world signals, where deviations from expected temporal structure can serve as early warning indicators of disruptive events such as cardiac arrhythmia, earthquakes, extreme weather or abrupt transitions in financial markets~\cite{lee_physical_2025, yang_seizure_2023, santos_reservoir_2025, walleshauser_predicting_2022, nadiga_reservoir_2021, yao_training_2015, noor_machine_2025}. In such systems, the dynamics are typically shaped by highly complex processes involving hidden degrees of freedom that are not directly observable and often give rise to pronounced memory effects, where the current state of the signal depends sensitively on its past. These characteristics make anomalies extremely difficult to identify. The MG series used above as a forecasting benchmark provides a prototypical example: although chaotic, it is generated by a delayed nonlinear differential equation that explicitly incorporates memory through a dependence on past values, and whose well-defined yet intricate algorithmic structure offers a meaningful reference against which deviations can be detected.

To illustrate the basic principle of anomaly detection using reservoir computing, we first introduce simple spiking perturbations into the MG signal (Fig.~\ref{fig:4}a). When a reservoir trained on unperturbed data processes this artificially altered signal, the prediction error exhibits sharp peaks exactly at the anomaly times (see Fig.~\ref{fig:4}b and Supplementary Movie~3), enabling straightforward identification by thresholding with an $\fscore$-score of 0.98 - a standard metric that quantifies classification performance. While instructive, this example is highly idealized and does not reflect realistic conditions, where anomalies are typically far more subtle and do not disrupt the signal. To emulate such scenarios, we therefore introduce a second example of “hidden" anomalies that preserve both instantaneous value and slope by skipping to a future matching segment of the signal (Fig.~\ref{fig:4}c). These perturbations leave the observable signal unchanged, including mean, variance, and short-time autocorrelation, yet disrupt the memory structure of the underlying process. As shown in Fig.~\ref{fig:4}d, these hidden anomalies nevertheless induce transient increases in prediction error, enabling reliable detection with an $\mathrm{F}_1$-score of 0.90. We want to emphasize that these results have been achieved in experiments and with $\Nhist = 1$, meaning the reservoir has no access to previous signal states but is entirely relying on its dynamic memory. Such direct hardware-level event extraction highlights the strong potential of active-matter reservoirs for edge-computing applications in which sensors deliver processed decisions rather than raw data streams. 

\section*{Outlook}
Our findings establish interacting ACOs as a powerful and fully accessible platform for physical reservoir computing, where hydrodynamic many-body coupling naturally provides the nonlinear dynamics and fading memory required for information processing. Unlike conservative interactions that depend solely on spatial separation, hydrodynamic forces depend on both distance and velocity, introducing time-asymmetric coupling and enabling dynamical responses that are difficult to realize in solid-state or fixed-network reservoirs. Crucially, all computations arise from the simultaneous dynamics of hundreds of coupled oscillators without any time-multiplexing so that the reservoir’s dimensionality directly reflects its genuine many-body behavior. Because lattice geometry, spatial arrangement, and subsets of driven oscillators can be freely programmed, the architecture can be precisely matched to the structure of incoming signals, offering exceptional flexibility for task-specific optimization. Beyond forecasting and anomaly detection, the high-dimensional dynamics generated by hydrodynamic coupling open opportunities for temporal-pattern classification, nonlinear filtering, closed-loop control, and autonomous microrobotic decision-making. Looking ahead, integrating adaptive feedback, programmable coupling, or viscoelastic media may unlock even richer dynamical modes and deeper memory, paving the way toward autonomous, energy-efficient edge-level computing architectures inspired by active matter.

\printbibliography

\newpage
\section*{Methods}\label{sec:methods}
\subsection*{Particle fabrication and propulsion}\label{meth:part_prep}
Active particles were manufactured by coating silica spheres (radius $a = 3\,\mathrm{\mu m}$) in one hemisphere with a $80\,\mathrm{nm}$ light-absorbing carbon layer. The particles were suspended in a mixture of water–2,6-lutidine (26.8 wt~\%), which exhibits a lower critical solution temperature at $34~^\circ\mathrm{C}$, and is confined in a quartz cell of $200\,\mathrm{\mu m}$ height held at $28~^\circ\mathrm{C}$. A focused $532\,\mathrm{nm}$ laser locally heats the carbon-coated side, inducing asymmetric demixing and thereby driving self-phoretic propulsion. The propulsion speed is controlled by the laser intensity, while the direction is defined by shifting the laser focus $2.4\,\mathrm{\mu m}$ away from the particle center opposite to the propulsion direction~\cite{gomez-solano_tuning_2017, lozano_phototaxis_2016}.

\subsection*{Particle steering and ACO realization}\label{meth:part_steering}
Laser steering was implemented by a two-axis acousto-optical deflector operated at 100~kHz, enabling quasi-simultaneous addressing of several hundred particles. Real-time feedback was obtained by continuously acquiring microscope images, extracting particle positions with sub-micrometer spatial and sub-second temporal resolution. An oscillatory motion of particle $i$ at position $\partpos_i$ is generated by continuously propelling it toward a spatially fixed target $\tarpos_i$. A feedback delay $\delay$ between particle detection and laser steering introduces a slight misalignment in the propulsion direction, producing a self-sustained orbital oscillation around the target site, i.e., an ACO (Figs.~\ref{fig:1}b–c)~\cite{heuthe_tunable_2025, chen_active_2023, wang_spontaneous_2023}.

\subsection*{Damping mechanism}\label{meth:damping}
To tune the responsiveness of ACOs to external perturbations, we implement a controlled damping scheme that modulates each particle’s propulsion velocity based on its instantaneous displacement from the target position. When the distance $|\partpos_i - \tarpos_i|$ drops below a threshold value $\rampLength$, the propulsion speed is reduced linearly with $|\partpos_i - \tarpos_i|$, while the propulsion direction remains unchanged. This mechanism does not stabilize the oscillation amplitude to a fixed value but provides continuous control over the extent of the oscillations around $\tarpos_i$ (Figs.~\ref{fig:1}e-f). We characterize the resulting dynamics through the z-component of the specific angular momentum of each oscillator $\angMom_i = \left(\vel_i \times (\partpos_i - \tarpos_i) \right) \cdot \hat{e}_z$. Without damping ($\rampLength = 0$), the ACOs dynamics are dominated by active oscillations with fluctuating amplitude (see Figs.~\ref{fig:1}e,g). As $\rampLength$ increases, these oscillations are progressively suppressed (Figs.~\ref{fig:1}f,h). The resulting changes are reflected in the narrowing of the distribution of $\left| \angMom_i \right|$ (Fig.~\ref{fig:1}i), demonstrating that $\rampLength$ provides a continuous control parameter for tuning the reservoir’s stability and noise robustness.

\subsection*{Memory time measurement}\label{meth:relaxation}
To probe the reservoir’s relaxation dynamics, we excite the ACO array with a single rectangular pulse implemented as a linear forward/backward shift of all target positions $\tarpos_i$ by $4\,\mathrm{\mu m}$ in the $x$-direction for a duration of $4\,\delay$. The resulting change in the mean particle velocity $\langle \scalarvel \rangle$ is recorded for different combinations of $\gridDist$ and damping parameter $\rampLength$. Strong hydrodynamic coupling at small $\gridDist$ produces large-amplitude collective oscillations, whereas increased spacing or higher damping progressively suppress these modes (see Figs.~\ref{fig:2}a-d). For a quantitative analysis, we compute the running variance $\moveVar{\langle \scalarvel \rangle}$ of the mean scalar velocity and fit its decay with a single exponential to extract the relaxation time $\tRelax$. These values reveal a pronounced, nontrivial dependence on both $\gridDist$ and $\rampLength$, spanning more than one order of magnitude (Fig.~\ref{fig:2}e).

\subsection*{Readout Training}\label{meth:readout_training}
After the data was collected and pre-processed from the experiment, the raw reservoir state trajectories were converted into readout trajectories constructed from Gaussian kernels.
The linear-, lasso-, and ridge-regression algorithms were used in the scikit-learn library~\cite{pedregosa_scikit-learn_2018}. The first 80\% of these trajectories were used for optimizing the weights of the readout layer $\readoutLayer$ using the ridge-regression algorithm in the scikit-learn library~\cite{pedregosa_scikit-learn_2018}. The remaining data was used to evaluate the prediction or anomaly detection performance.

\subsection*{Reservoir Computing}\label{meth:reservoir-computing}
A reservoir computer can be divided into two components: the reservoir and the readout layer.
The reservoir, $f$, driven by an input signal, $\signal(t) \in \mathbb{R}^{N}$, produces a high-dimensional representation, $\resState(t)$, referred to as a feature vector, such that
\begin{equation}
    \centering
    f : \mathbb{R}^{N} \to \mathbb{R}^{M}, \quad \signal(t) \in \mathbb{R}^{N} \mapsto \resState(t) \in \mathbb{R}^{M}
\end{equation}
with $M \gg N$.
The key to a useful reservoir is the amount of information captured by these representations $\resState(t)$.
Ideally, this mapping is such that similar values in the input space, defined by a trajectory, $\tau$, are mapped to similar representations in the reservoir embedding, that is, $\langle\signal_{1}(\tau)\cdot\signal_{2}(\tau)\rangle \approx 1 \implies \langle\resState{1}(\tau)\cdot \resState{2}(\tau)\rangle \approx 1$, and distinct values are mapped far apart, for well-defined inner products.
Naturally, this introduces a balance between the amount of nonlinearity one needs and the rich structure that will be required to perform computing tasks using the feature vectors.
In the reservoirs studied here, a system of coupled particles is used as a reservoir, thus the state of the system after being driven can be described by a coordinate in phase space, $(\{\partpos_{i}\}, \{\vel_{i}\}) \in \mathbb{R}^{2N}$ where $\{\partpos_{i}\} = \{\partpos_{0}, \partpos_{1}, \dots, \partpos_{N}\}$ are the positions of the $N$ particles in the reservoir, and $\{\bm{v}_{i}\} = \{\vel_{0}, \vel_{1}, \dots, \vel_{N}\}$ are their velocities.
We do not consider momentum as we are dealing with active particles in a low Reynolds number regime.

The second component in reservoir computing is the readout layer.
Once the signal $\signal(t)$ has driven the reservoir into some state $\resState(t)$, a measurement can be performed to construct a feature vector $\readout(t)$ describing that state, on which a readout layer with weights $\readoutLayer$ is applied to produce a prediction $\resOutput(t+\predLen)$ based on the task being performed:
\begin{equation}
    \resOutput(t + \predLen) = \readoutLayer \readout(t)^\top.
\end{equation}
The construction of this feature vector $\readout$ is the subject of Section "Reservoir Measurements" as it constitutes an open problem in reservoir computing, the method by which this vector is turned into a prediction, however, is far better defined.
As the complex, nonlinear part of the task has been handled by the reservoir, these feature vectors are typically passed through a simple regression algorithm such as ridge regression lasso regression, or even plain linear regression without any regularization to optimize the weights of $\readoutLayer$.
It is also common that a single linear neural network layer is trained for the readout by minimizing a loss function related to the task.

\subsection*{Reservoir Measurements}\label{meth:reservoir-measurements}
As mentioned in Methods section "Reservoir Computing", constructing a feature vector that describes the state of a reservoir at any moment in time is an open problem in the field.
In the case of a reservoir of active particles, a simple approach might be the positions and velocities of each particle in the system, a complete description of the phase space coordinate.
However, as the system scales, this vector will become highly overcomplete and contain what might be a large degree of irrelevant information, thus making the readout slower and more challenging.

In this study, the Gaussian kernel approach introduced in~\cite{lymburn_reservoir_2021} has been applied due to its simplicity and effectiveness.
When using the Gaussian kernels, $m$ Gaussian functions of the form
\begin{equation}
    \centering
    \Psi(\bm{x}) = \exp\left[-{\frac{\left(\bm{x} - \bm{c}_{m}\right)^2}{2w_{m}^{2}}}\right],
\end{equation}
where $\bm{c}_{m}$ is the spatial coordinate of the center kernel and $w_{m}$ is the width, are placed around the system and are used to compute three observables:
\begin{align}
\centering
&o_{1, m}(t) = \sum\limits_{i=1}^{N}\Psi_{m}(x_{i}(t)) \\
&o_{x, m}(t) = \sum\limits_{i=1}^{N}\Psi_{m}(x_{i}(t))v_{xi}(t)
\end{align}
where $N$ sums over all particle considered in the receptive field of the kernel, and the $x$ subscript in the second observable can be replaced by the number of unique velocity vectors available, in our case, $x, y$, resulting in three total observables.
The width of the kernel, $w_{m}$ is chosen based on the distance to the 5th nearest neighbor from the chosen observation point, which is typically chosen as a random particle position. 
In this way, in regions where there are lots of particles, the Gaussian kernel is less smeared and contains more dense information.

\subsection*{Chaotic time series generation}
The chaotic Mackey-Glass time series was generated with the Signalz python package and parameters $a = 0.2$,  $b = 0.8$, $c=0.9$, $e=10$ and an initial value of 0.1.
The chaotic Lorenz attractor trajectory was taken from the datasets available in the reservoirpy python package, which is the Lorenz63 system with parameters $\rho = 28$, $\sigma = 10$, $\beta = 2.66$, $h = 0.03$ and initial values of 1 for $x$, $y$ and $z$ coordinates.
Both of these time series are well established benchmarks for chaotic signal forecasting~\cite{wringe_reservoir_2025}.

\subsection*{Hydrodynamic Coupling Model}\label{meth:hydrodynamic-interaction-theory}
The interaction of the ACOs in our study is dominated by hydrodynamic interactions. In particular, in the case of active particles, such interactions can be rather complex, due to the three-dimensional structure of the surface flows generated by the propulsion mechanism. At the particle distances covered in our experiments, hydrodynamic interaction between active particles can be approximated with a Kuramoto-type coupling~\cite{uchida_synchronization_2010}.
This approximation, however, is not sufficient to capture the conditions in our experiments. Therefore, we employ a more detailed hydrodynamic interaction model, which has been shown to yield good agreement with our experiments and which is explained below in brief (for a detailed description, we refer to ~\cite{heuthe_tunable_2025}).
When an active colloidal particle moves, it generates a flow field $\mathbf{U}_{\text{HI}}(\partpos)$ within the surrounding fluid that exerts hydrodynamic forces on nearby particles. In our system
\begin{equation}
    \mathbf{U}_{\text{HI}}(\partpos) = \mathbf{U}_{flow}(\partpos) +  \mathbf{U}_{lub}(\partpos).
\end{equation}
where $\mathbf{U}_{flow}(\partpos)$ results from the propulsion mechanism of the active colloid and $\mathbf{U}_{lub}(\partpos)$ is the lubrication flow field  that emerges from the fluid being displaced between the moving colloid and another particle. $\mathbf{U}_{flow,i}(\partpos)$ around colloid $i$ can be approximated with a so-called squirmer model~\cite{lighthill_squirming_1952, blake_spherical_1971}, which has the form:
\begin{equation}
     \mathbf{U}_{flow,i}(\partpos) = -\frac{p}{r^2}[1-3(\mathbf{e}_i\cdot\mathbf{\hat{r}})^2]\mathbf{\hat{r}} - \frac{s}{r^3}[\mathbf{e}-3(\mathbf{e}_i\cdot\mathbf{\hat{r}})\mathbf{\hat{r}}]
\end{equation}
where $\mathbf{\hat{r}}=\partpos/\left| \partpos \right|$.
The fluid field to model the lubrication interactions between two colloids with radius $a$ and surface-surface distance $h$ can be written as
\begin{equation}
    \mathbf{U}_{lub}(\partpos) = -\frac{a}{h} (\vel\cdot \mathbf{\hat{r}}) \Theta(r) \mathbf{\hat{r}}.
\end{equation}
To obtain the total force exerted on particle $j$ at position $\partpos_j$, we sum up the flow fields of all other colloids $i$ around $j$
\begin{equation}
    \mathbf{U}_{\text{HI}}(\partpos_i) = \sum_i \mathbf{U}_{flow,i}(\partpos_j-\partpos_i) +  \mathbf{U}_{lub,i}(\partpos_j-\partpos_i).
\end{equation}
From this combined flow field, we compute the force acting on colloid $j$ using
\begin{equation}
    \mathbf{F}_{\text{HI}}(\partpos_j)=\gamma \mathbf{U}_{\text{HI}}(\partpos_j).
\end{equation}

\section*{Data availability}
The data that support the findings of this study are available at Zenodo (DOI: 10.5281/zenodo.17913226).

\section*{Code availability}
The code used to generate and analyze the data of this study is available at Zenodo (DOI: 10.5281/zenodo.17913226).

\section*{Acknowledgments}
We acknowledge Priyanka Iyer for her help with the simulation code and Christian Holm for discussion on the manuscript. C.B. acknowledges funding by the ERC through the Adv. Grant BRONEB (101141477). V.L.H. and C.B. acknowledge funding from the DFG Centre of Excellence 2117, Germany, Centre for the Advanced Study of Collective Behavior (ID:422037984).

\section*{Author contributions}
V.L.H., S.T. and C.B. devised the research. V.L.H. performed the experiments, wrote evaluation code, analyzed data and prepared the figures. L.S. performed the simulations, wrote evaluation code and analyzed data. S.T. wrote evaluation code. V.L.H., S.T. and C.B. wrote the paper.

\section*{Competing interests}
The authors declare no conflict of interest.

\section*{Additional Information}
Supplementary Information is available for this paper.
Correspondence and requests for materials should be addressed to C.B. (clemens.bechinger@uni.kn).

\clearpage
\input{preprint/SI_preprint}
\end{document}

%% file: math_commands.tex
\usepackage{amsmath,amsfonts,bm}

\def\eqref#1{equation~\ref{#1}}

\def\1{\bm{1}}

\DeclareMathAlphabet{\mathsfit}{\encodingdefault}{\sfdefault}{m}{sl}
\SetMathAlphabet{\mathsfit}{bold}{\encodingdefault}{\sfdefault}{bx}{n}



%% file: preprint/SI_preprint.tex
\renewcommand{\predError}{\epsilon}
\renewcommand{\resOutput}{\mathbf{o}}
\renewcommand{\delay}{\tau_{\text{d}}}
\renewcommand{\vel}{\mathbf{v}}
\renewcommand{\scalarvel}{v}
\renewcommand{\partpos}{\mathbf{r}}
\renewcommand{\rampLength}{\Gamma}
\renewcommand{\laserInt}{I}
\renewcommand{\Izero}{I_{\textnormal{0}}}
\renewcommand{\tarDist}{\mathbf{d}_{\textnormal{t}}}
\renewcommand{\tarDisti}{\mathbf{d}_{\textnormal{t,i}}}
\renewcommand{\abstarDist}{D_{\textnormal{target}}}
\renewcommand{\angMom}{h}
\renewcommand{\tarpos}{\mathbf{q}}
\renewcommand{\tarMotAng}{\beta}
\renewcommand{\gridDist}{L}
\renewcommand{\tRelax}{\tau_{r}}
\renewcommand{\meanAbsVel}{\langle \scalarvel \rangle}
\renewcommand{\moveVar}[1]{\sigma^2_{10}[#1]}
\renewcommand{\signal}{\mathbf{u}}
\renewcommand{\resState}{\mathbf{s}}
\renewcommand{\readout}{\mathbf{x}}
\renewcommand{\predLen}{\Delta t}
\renewcommand{\coeffDet}{\mathrm{R}2}
\renewcommand{\driverFract}{f_{\mathrm{driver}}}
\renewcommand{\Nhist}{N}
\renewcommand{\readoutLayer}{\mathbb{W}_\mathrm{out}}
\renewcommand{\timeComp}{f_c}
\renewcommand{\nrmse}{N\!R\!M\!S\!E}
\renewcommand{\fscore}{F_{1}}

\onecolumn
\begin{center}
{\LARGE \bfseries Supplementary Information to Reservoir computing from collective dynamics of active colloidal oscillators\par}
\vspace{1em}
\normalsize
Veit-Lorenz Heuthe$^{1,2}$, Lukas Seemann$^{1}$, Samuel Tovey$^{3}$, and Clemens Bechinger$^{1, 2, \dag}$ \par
\vspace{0.5em}
\small
$^1$Fachbereich Physik, Universität Konstanz, D-78465 Konstanz, Germany \\
$^2$Centre for the Advanced Study of Collective Behaviour, Universität Konstanz, 78464 Konstanz, Germany \\
$^3$Institute for Computational Physics, University of Stuttgart, 70569 Stuttgart, Germany \par
\vspace{0.5em}
$^\dag$Corresponding author. Email: clemens.bechinger@uni-konstanz.de\par
\end{center}

\begin{suppfig}[H]
    \centering
    \includegraphics[width=1.0\linewidth]{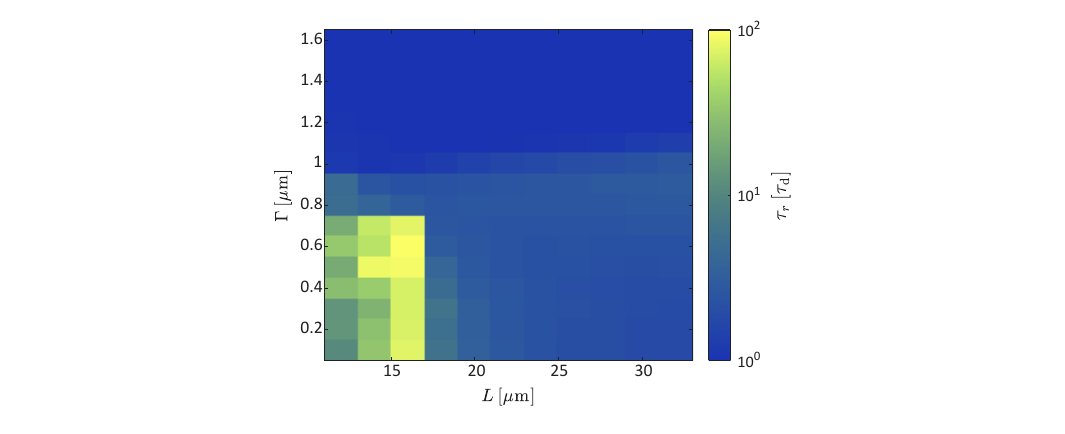}
    \refstepcounter{sifig}
    \caption{
        Relaxation of the reservoir in simulations with arrays of 400 ACOs after excitation by displacing all target positions in the x-direction by $4\,\mathrm{\mu m}$ for $4\,\mathrm{\tau_d}$.
        }
    \label{si_fig:sim_relax_times}
\end{suppfig}

\begin{suppfig}[H]
    \centering
    \includegraphics[width=1.0\linewidth]{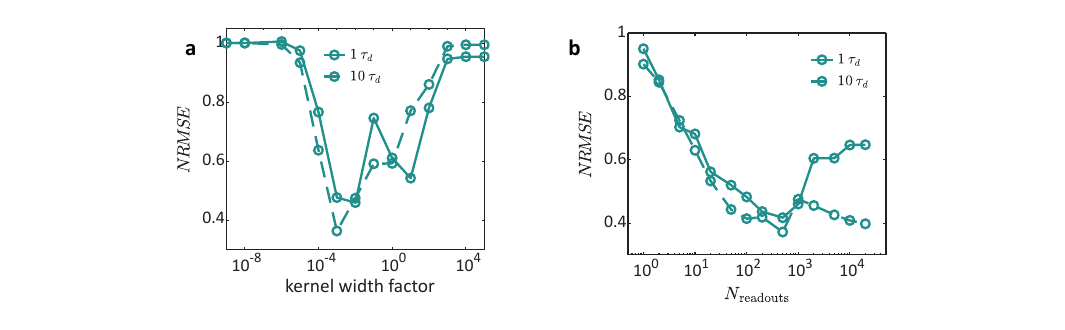}
    \refstepcounter{sifig}
    \caption{
        \bf{Optimization of Gaussian kernel width and number for the readout.}
        }
    \label{si_fig:kernel_optimization}
\end{suppfig}

\begin{suppfig}[H]
    \centering
    \refstepcounter{sifig}
    \caption{
        \textbf{MG forecasting errors in experiment}
        \textbf{a} Normalized root mean squared error $\nrmse$ in experiments as a function of the number $\Nhist$ of previous signal states that are fed in to the reservoir.
        \textbf{b} $\nrmse$ against the number $\Nhist$ of injected states in the reservoir in simulation (extension to Fig.~3f in the main text).
        \textbf{c} $\nrmse$ as a function of the fraction $\driverFract$ of ACOs that receive the input signal in experiments.
        }
    \includegraphics[width=1.0\linewidth]{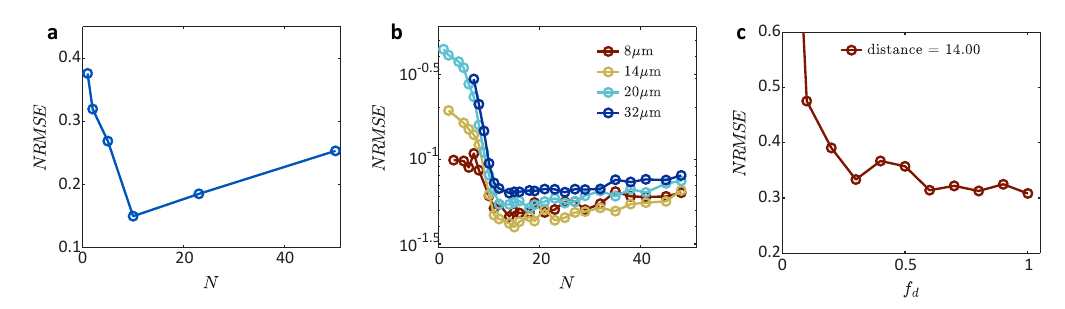}
    \label{si_fig:exp_MG_pred_ranges}
\end{suppfig}

\section*{Chaotic Lorenz attractor trajectory forecasting} \label{si:lorenz_pred}
\begin{suppfig}[H]
    \centering
    \includegraphics[width=1.0\linewidth]{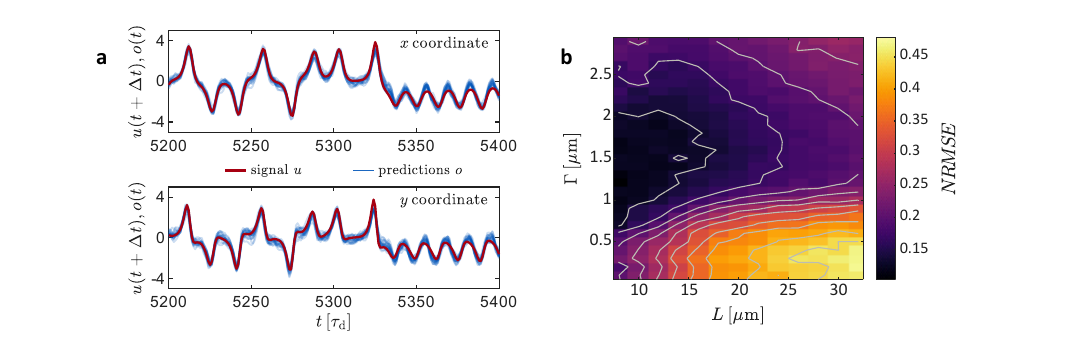}
    \refstepcounter{sifig}
    \caption{
        \textbf{Forecasting of the chaotic Lorenz63 attractor}
        \textbf{a} Signal (red) and predictions (blue) for x- and y-coordinate (top and bottom, respectively) of the chaotic Lorenz63 attractor in experiments.
        \textbf{b} $\nrmse$ for forecasting the x-coordinate of the Lorenz63 system with $\predLen = 2\delay$ in simulations as a function of the distance $\gridDist$ between ACOs and damping parameter $\rampLength$.}
    \label{si_fig:lorenz_pred}
\end{suppfig}

In addition to the chaotic Mackey-Glass (MG) time series forecasting benchmark presented in the main text (Fig.~3), we have tested the ACO reservoir in predicting the Lorenz63 differential equation system, which is another well established reservoir computing benchmark (for parameters see Methods).
Figure~\ref{si_fig:lorenz_pred}a shows predictions of the x- and y coordinate from experiments, where the target position of each ACO was driven with the two-dimensional x-y signal of the Lorenz63 attractor.
Figure~\ref{si_fig:lorenz_pred}b shows the $\nrmse$ for predicting the x-coordinate of the Lorenz63 system in simulations as a function of the separation $\gridDist$ between the ACOs in the array and the damping parameter $\rampLength$ for $\predLen = 2\delay$.
Qualitatively, the prediction performance shows the same dependence on these two parameters as for the MG forecasting task (compare to Fig.~3d in the main text), which shows that the observed effects are independent of the task but depend on the properties of the reservoir.